\documentclass[twocolumn,aps,pre,superscriptaddress]{revtex4}
\usepackage{graphicx}
\usepackage{dcolumn}
\usepackage{amsmath}
\usepackage{latexsym}

\begin {document}

   \title {Jamming and percolation properties of random sequential adsorption with relaxation}

   \author{Sumanta Kundu}
   \email[ ]{sumanta492@gmail.com}
   \affiliation{
   \begin {tabular}{c}
   Satyendra Nath Bose National Centre for Basic Sciences, Block-JD, Sector-III, Salt Lake, Kolkata-700106, India
   \end {tabular}
   }

   \author{Nuno A.~M.~Ara\'{u}jo}
   \email[ ]{nmaraujo@fc.ul.pt}
   \affiliation{
   \begin {tabular}{c}
   Departamento de F\'{\i}sica, Faculdade de Ci\^{e}ncias, Universidade de Lisboa, 1749-016 Lisboa, Portugal
   \end {tabular}
   }

   \author{S.~S.~Manna}
   \email[ ]{subhrangshu.manna@gmail.com}
   \affiliation{
   \begin {tabular}{c}
   Satyendra Nath Bose National Centre for Basic Sciences, Block-JD, Sector-III, Salt Lake, Kolkata-700106, India
   \end {tabular}
   }

\begin{abstract}
 
      The random sequential adsorption (RSA) model is a classical model in Statistical Physics for 
   adsorption on two-dimensional surfaces. Objects are deposited sequentially at random and adsorb 
   irreversibly on the landing site, provided that they do not overlap any previously adsorbed 
   object. The kinetics of adsorption ceases when no more objects can be adsorbed (jamming state). 
   Here, we investigate the role of post-relaxation on the jamming state and percolation properties 
   of RSA of dimers on a two-dimensional lattice. We consider that, if the deposited dimer partially 
   overlaps with a previously adsorbed one, a sequence of dimer displacements may occur to accommodate 
   the new dimer. The introduction of this simple relaxation dynamics leads to a more dense jamming 
   state than the one obtained with RSA without relaxation. We also consider the anisotropic case, 
   where one dimer orientation is favored over the other, finding a non-monotonic dependence of the 
   jamming coverage on the strength of anisotropy. We find that the density of adsorbed dimers at 
   which percolation occurs is reduced with relaxation, but the value depends on the strength of 
   anisotropy.
 
\end{abstract}

\maketitle

\section{Introduction}

      Adsorption of geometrical objects on a substrate has been a problem of great interest due to its
   applicability in a variety of fields ranging from photonic crystals to quantum dots including, e.g., 
   surface coating and encapsulation~\cite{Privman,Kumacheva,Lewis,O'Conor,Evans,Torquato,Burda,Oliveira}. 
   Theoretically, the model of Random Sequential Adsorption (RSA) has been studied intensively over last 
   decades in the context of irreversible processes of adsorption on surfaces~\cite{Bartelt,Privman2,Cadilhe}.
   P. J. Flory introduced the RSA model in a one-dimensional chain to study the interaction between blocks 
   along a linear polymer chain~\cite{Flory}. This model attracted great attention from the scientific 
   community and was later interpreted as a problem of critical phenomena by R\'{e}nyi~\cite{Renyi} and 
   Feder~\cite{Feder}.

      In RSA, the objects are adsorbed sequentially and irreversibly at randomly selected vacant positions 
   on a surface. Selection of occupied positions are discarded due to the excluded volume interaction with 
   the previously adsorbed objects. These objects are assumed to be inherently immobile, i.e., they never 
   move out from their positions after adsorption. The interesting feature of this model is the existence 
   of a non-trivial jamming state where no more objects can be adsorbed~\cite{Manna,Evans}.

      Subsequently, a number of variants of the RSA model have been studied to explain the observations
   of various natural and experimental scenarios~\cite{Evans,King,Guo,Rodgers,Ciesla,Joshi,Pinto}. For
   example, the model of accelerated RSA was introduced to describe the mechanism of precursor mediated
   chemisorption~\cite{Guo}. In this model, if the deposited object lands on top of the already adsorbed
   objects it starts diffusing till it finds a vacant gap where it is adsorbed irreversibly~\cite{Rodgers}.

      The configuration of objects at any arbitrary intermediate stage of the RSA process corresponds to
   a disordered system and the study of their percolation properties~\cite{Broadbent,Stauffer,Grimmett,
   Sahimi,Sahimi2,Araujo,Saberi,Lee} is of interest. To describe briefly, in percolation the sites (bonds) 
   of a regular lattice are occupied with probability $p$ or kept vacant with probability $(1-p)$. These 
   occupied sites (bonds) form clusters of different sizes through their neighboring connections. A 
   continuous transition between the ordered and disordered phases is observed at a critical value of 
   $p=p_c$. For $p > p_c$, there exists global connectivity through macroscopic cluster that scales 
   linearly with the volume of the system. Numerically, this usually the one that spans between two opposite 
   sides of the lattice~\cite{Stauffer}. Till date, the best value of $p_c$ for the site percolation on the 
   square lattice is $0.59274605079210(2)$~\cite{Jacobsen} and exactly $1/2$ for the bond percolation.

      In this paper, we introduce a variant of RSA where the objects (dimers) are adsorbed irreversibly 
   onto the lattice sites after going through a well defined relaxation dynamics. We consider a very 
   simple relaxation dynamics where, during the relaxation, a series of dimer displacements may occur 
   to accommodate the new dimer. The effect of such a relaxation dynamics and anisotropy in the 
   orientation of the adsorbed dimers on the jamming state and percolation transition are investigated 
   here using numerical simulations.

\section{Model}

      Dimers are adsorbed sequentially at random positions onto an initially empty square lattice of size
   $L \times L$, with periodic boundary condition. Each dimer occupies two lattice sites. To attempt the 
   adsorption of a dimer, its orientation (either vertical or horizontal) is first selected randomly with 
   equal probability for both orientations. A pair of neighboring sites are then selected accordingly at 
   random and the dimer is deposited on them.

\begin{figure}[t]
\begin{center}
\begin {tabular}{cc}
\includegraphics[width=0.5\linewidth]{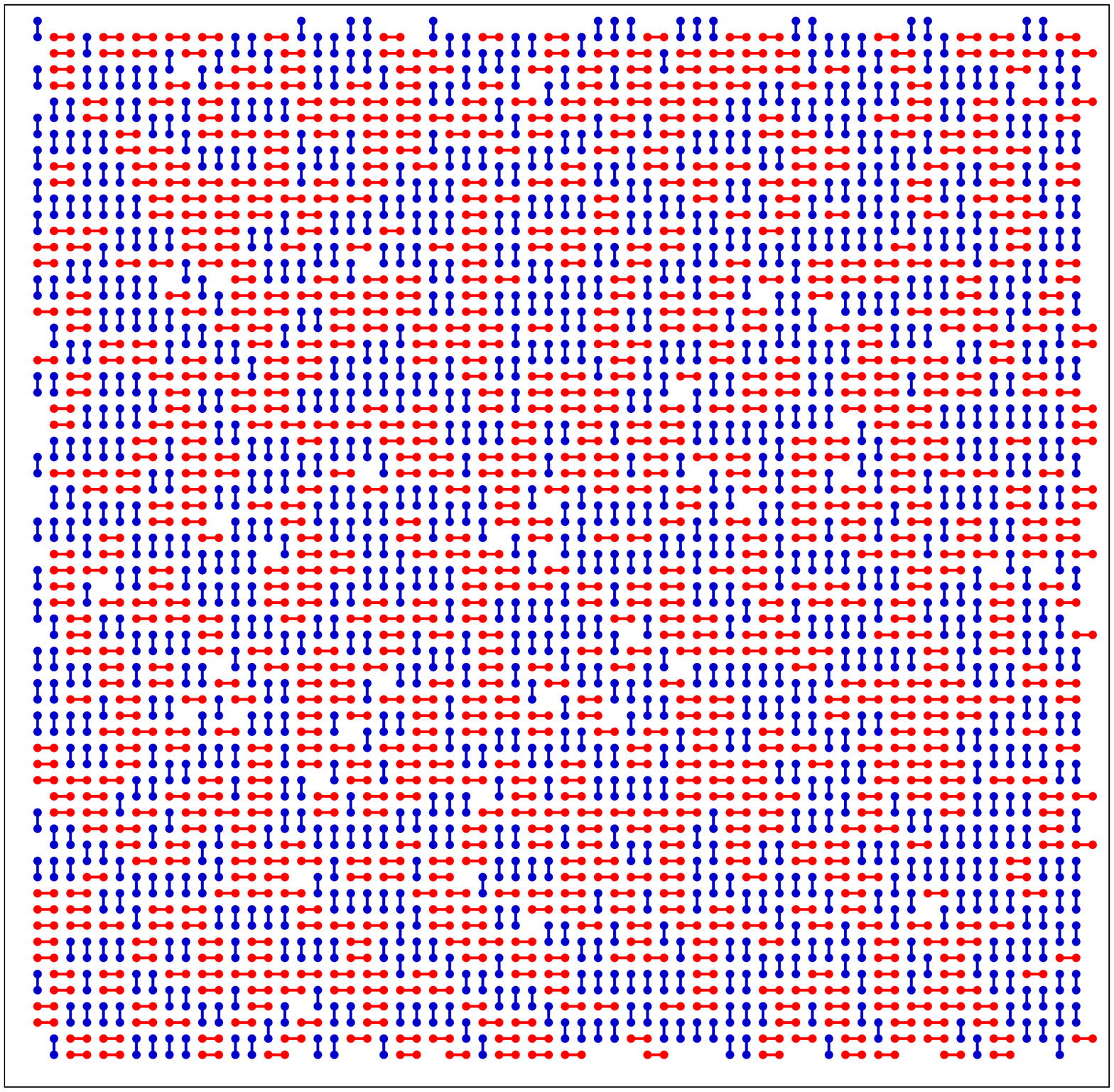} & \includegraphics[width=0.5\linewidth]{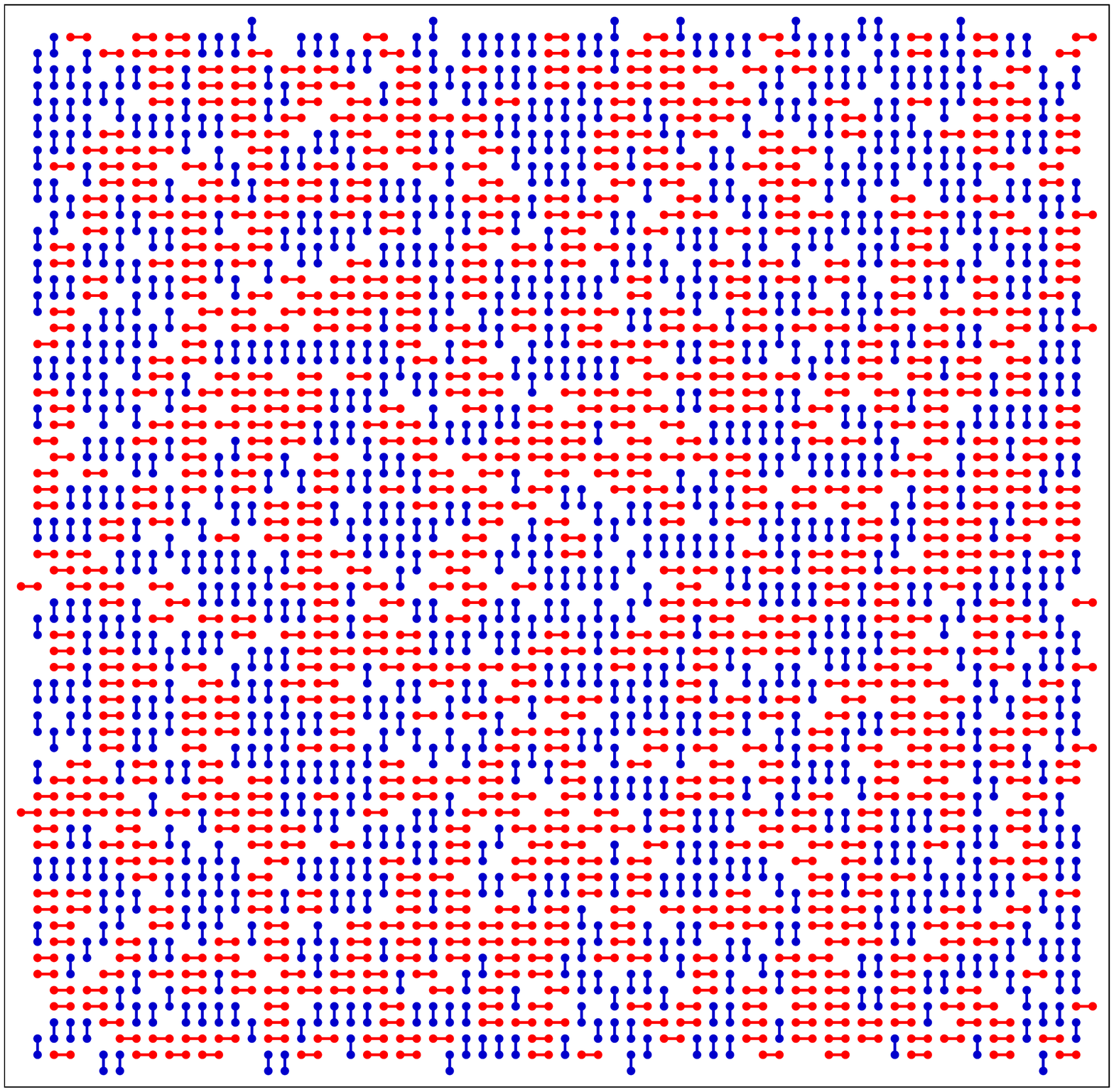}\\
(a) & (b)
\end {tabular}
\end{center}
\caption{Typical jamming state configuration of the dimers on a 64 $\times$ 64 square lattice for the Random 
   Sequential Adsorption model (a) with and (b) without relaxation. The dimers oriented in the horizontal and 
   vertical directions have been painted in red and blue colors, respectively. The single vacant sites are 
   represented by white color. 
}
\label{JAMPC}
\end{figure}

      Depending on the occupation state of the pair of sites, there are three possible outputs. First,
   if the pair of sites are both occupied by previously adsorbed dimers, adsorption fails. Second, if 
   both sites are vacant, the adsorption is successful and the dimer is irreversibly adsorbed on them. 
   Third, if only one of the sites is vacant, a sequence of dimer displacements is triggered, described 
   as follows. When the deposited dimer (A) overlaps with a previously adsorbed dimer (B) at one end, 
   the dimer B is displaced by a unit distance along its other end, keeping A fixed. The displaced dimer 
   may partially overlap with another dimer (C) leading to similar displacement of C. The system of 
   adsorbed dimers thus relaxes and eventually reaches a stable state when no more overlapping of dimers 
   exists. This concerted move completes the ``successful'' adsorption of dimer A through a relaxation 
   process. Here one assumes the existence of two infinitely separated time scales, as we consider that 
   the relaxation process is always faster than the inter-arrival time of deposited particles. The trail 
   of dimer displacements originated by depositing A constitute a path which is referred as the 
   ``relaxation path''. It has been observed that often a relaxation path forms a closed loop. In such a 
   case, the deposition attempt fails and the deposited dimer is discarded. The sequence of dimer adsorption 
   attempts is continued till a jamming state is reached, where no more dimers can adsorb.

\begin{figure}[t]
\begin{center}
\includegraphics[width=\linewidth]{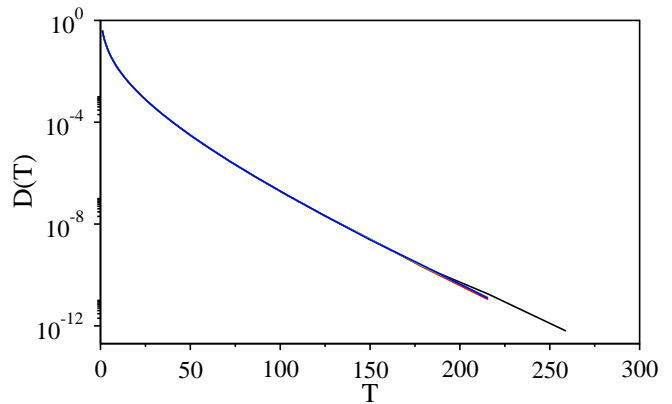}
\end{center}
\caption{Plot of the binned data of the relaxation time distribution $D(T)$ for the entire process of 
   adsorption on a log-lin scale using the lattice sizes $L = 256$ (black), $512$ (red), $1024$ (green), 
   and $2048$ (blue). The data points are averages over samples ranging from $2 \times 10^6$ for $L=256$ 
   to 12000 for $L=2048$.
}
\label{RTIME}
\end{figure}

      The coverage of the surface is defined as $p=2n/L^2$, where $n$ is the number of adsorbed dimers.
   An occupied site can never become vacant, since there are no desorption events during the relaxation.
   When $p$ is small, adsorption of dimers is mainly uncorrelated and post-relaxation is negligible. For
   intermediate values of $p$, successful adsorptions are often associated with relaxation. In this case, 
   the newly occupied pair of sites are positioned at the two ends of the relaxation path and are separated 
   by a distance larger than unity. Thus, the relaxation process introduces spatial correlations between 
   occupied sites. Such a source of correlation is absent in the model of RSA without relaxation. By further 
   increasing the value of $p$, the clusters of occupied sites start merging leading to a percolation 
   transition. The percolation threshold $p_c$ is defined as the minimum value of $p$ for which a giant 
   cluster emerges that spans the entire surface, touching opposite ends of the lattice. This percolation 
   transition is observed before the jamming transition.

\section{Results}
\subsection{Jamming state and relaxation time}

      The averaged fraction of the occupied sites at the jamming state defines the jamming coverage $p_j$. 
   Figure \ref{JAMPC}(a) depicts a typical jamming state configuration of RSA model with relaxation and we 
   compare it with one obtained for RSA without relaxation, Fig.\ \ref{JAMPC}(b). The relaxation dynamics 
   promotes the reorganization and packing of the dimers more densely so that the jamming state coverage 
   is larger than that of the RSA without relaxation. Numerically, we have estimated the jamming state 
   coverage $p_j(L)$ and its standard deviation $\Delta(L) = (\langle p_j^2 \rangle - \langle p_j \rangle ^2)^{1/2}$ 
   for different system sizes $L$ = 256, 512, 1024, 2048, and 4096. We observe no significant finite-size 
   effects for $p_j(L)$ and that its value is $0.99049(3)$ compared to $0.90682(3)$ for RSA without 
   relaxation. The value of $\Delta(L)$ indeed varies significantly with $L$. Fitting to a power-law decay: 
   $\Delta(L) \sim L^{-1/\nu_j}$ we have estimated $\nu_j$ = 1.002(3), consistent with a linear decay with 
   $1/L$.

      The duration of the relaxation process triggered by the deposition of a dimer is termed as the 
   ``relaxation time'' $T$ and it corresponds to the number of successive dimer displacements before 
   a successful adsorption. This relaxation time has been measured for every dimer deposited from the 
   beginning till the jamming state and its distribution $D(T)$ is plotted for four different system 
   sizes in Fig.\ \ref{RTIME}. Clearly, the tail of the distribution decays exponentially in time 
   suggesting a characteristic time of $\approx$ 11.7, in units of dimer displacements.

\begin{figure}[t]
\begin{center}
\includegraphics[width=\linewidth]{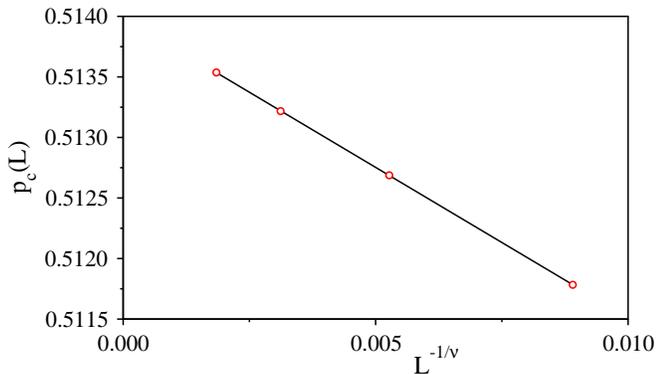}
\end{center}
\caption{Plot of the percolation threshold $p_c(L)$ against $L^{-1/\nu}$ with $1/\nu = 0.756(6)$ for the lattice
   sizes $L = 512, 1024, 2048,$ and $4096$. The asymptotic value of the percolation threshold in the limit $L \to 
   \infty$ has been estimated to be 0.5140(1). The data points are averages over samples ranging from $9 \times 
   10^5$ for $L=512$ to 8900 for $L=4096$.
}
\label{pc}
\end{figure}

\subsection{Percolation transition}

      As the surface coverage $p$ increases, the size of the largest cluster grows monotonically. Numerically, 
   the precise value of the percolation threshold $p_c^\alpha$ for a specific run $\alpha$ is determined using 
   the bisection method~\cite{Kundu} described as follows. We select a pair of initial values of $p$, namely, 
   $p^{\text{h}}$ and $p^{\text{l}}$ such that there exists a global connectivity through the spanning cluster 
   for $p=p^{\text{h}}$ but not for $p^{\text{l}}$. Starting from an empty lattice the adsorption is continued 
   till the density of occupied sites $p = (p^{\text{h}} + p^{\text{l}})/2$ is reached. Here, connectivity 
   between the top and the bottom sides of the lattice is checked using the burning algorithm~\cite{Stauffer} 
   while imposing periodic boundary condition along the horizontal direction. If the opposite sides of the 
   lattice are connected by the same cluster, $p^{\text{h}}$ is reduced to $p$, otherwise $p^{\text{l}}$ is 
   raised to $p$. In this way, the interval is iteratively bisected until $p^{\text{h}} - p^{\text{l}} < 2/L^2$, 
   when $(p^{\text{h}} + p^{\text{l}})/2$ defines the value of $p_c^\alpha$. The entire procedure is then 
   repeated for a large number of independent runs and the individual percolation thresholds are averaged to 
   obtain the estimated percolation threshold $p_c(L) = \langle p_c^\alpha (L) \rangle$ for the surface of size 
   $L$. These values are then extrapolated to obtain the asymptotic value $p_c$ in the limit $L \to \infty$ using,
\begin{equation}
p_c(L) = p_c - AL^{-1/\nu},
\label{EQN01}
\end{equation}
   where $\nu$ is known as the correlation length exponent in percolation theory and its value is $4/3$ for 
   random percolation in two dimensions~\cite{Eschbach,Stauffer}. The obtained values of $p_c(L)$ are plotted 
   against $L^{-1/\nu}$ in Fig.\ \ref{pc}. Tuning the value of $1/\nu$, the data is found to be fit best by a 
   straight line (using the least square fit of a straight line with minimal error) for $1/\nu$ = 0.756. By
   extrapolating to the thermodynamic limit ($L \to \infty$), we obtain $p_c$ = 0.5140(1). This value is much 
   smaller than the value of $p_c$ = 0.5618(1) for the RSA without relaxation~\cite{Vandewalle, Cherkasova}. 

      Qualitatively, one can try to understand the reduction in the value of percolation threshold in the 
   following way. Let us consider a situation where a single vacant site P separates two distinct clusters 
   connected to the top and bottom boundaries. In RSA without relaxation, it needs a dimer to be adsorbed
   precisely on this vacant site to connect the two. With relaxation, a dimer may be deposited at many other 
   locations, yet due to the relaxation process another dimer may be displaced to the site P and connect 
   the two clusters. 

\begin{figure}[t]
\begin{center}
\begin{tabular}{c}
\includegraphics[width=\linewidth]{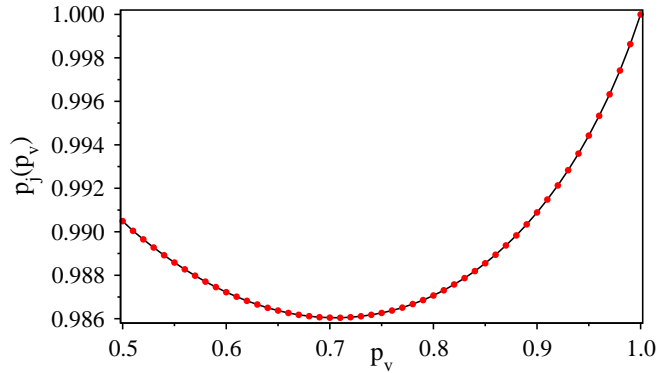}
\end{tabular}
\end{center}
\caption{For $L$ = 1024, the jamming state coverage $p_j(p_v)$ has been plotted against the selection probability 
   $p_v$ of the vertically oriented dimers. The data points are averages over (at least) $10^5$ samples.}
\label{fig:pjpv}
\end{figure}

      To investigate the critical properties of the percolation transition of RSA with relaxation, several critical 
   exponents have been estimated. Using extensive numerical simulations, at $p=p_c$, we have determined the fractal 
   dimension of the largest cluster $d_f=1.892(2)$, the exponent $\gamma/\nu=1.790(2)$ associated with the second 
   moment of the cluster size distribution and the fractal dimension of the shortest path $d_l=1.1307(5)$. These 
   values are consistent, within error bars, with the values known for random percolation in two dimensions, namely, 
   $d_f=91/48$, $\gamma/\nu=43/24$~\cite{Stauffer} and $d_l=1.13077(2)$~\cite{Zhou}.

\begin{figure}[t]
\begin{center}
\begin{tabular}{c}
\includegraphics[width=\linewidth]{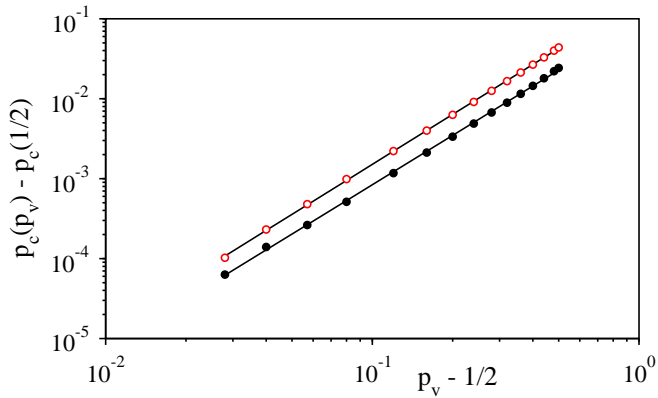}
\end{tabular}
\end{center}
\caption{Plot of the deviation of the percolation threshold $p_c(p_v) - p_c(1/2)$ against $p_v - 1/2$, $p_v$ 
   being the selection probability of the vertical dimers, on a log-log scale with $p_c(1/2)=0.5140(1)$ and 
   $0.5619(1)$ for the RSA with (open circles) and without (filled circles) relaxation, respectively. For 
   each value of $p_v$, the $p_c(p_v)$ in the limit of $L \to \infty$ has been obtained using the values of 
   $p_c(p_v,L)$ for $L=256, 512, 1024, 2048,$ and $4096$ and an extrapolation given by Eq.\ (\ref{EQN01}). 
   The slopes of fitted straight lines have been measured as $2.05(6)$ and $2.07(7)$, respectively. The data 
   points are averages over samples ranging from (at least) $1.6 \times 10^6$ for $L=256$ to 5000 for $L=4096$.
}
\label{fig:pcpv}
\end{figure}

\subsection{Effect of anisotropy on jamming and percolation}

      So far, we have considered that the orientation of the depositing dimers is drawn at random with equal 
   probability for horizontally and vertically oriented dimers. We now consider the anisotropic case, where 
   these probabilities are different. More specifically, when the $n$-th dimer is deposited, its orientation 
   is randomly selected with probability $p_v$ or $1 - p_v$ for vertical and horizontal, respectively. If 
   the deposition attempt fails, another dimer is deposited with the same orientation but at another location 
   (selected at random) until the adsorption is successful.

      For $p_v>1/2$, we observe that the clusters are elongated along the vertical direction. For this regime, 
   the jamming state is defined as a configuration where no more vertically oriented dimer can adsorb. It 
   turned out that the anisotropy affects significantly the value of the jamming state coverage, as shown in 
   Fig.\ \ref{fig:pjpv}, with $p_j=0.99049$ for $p_v=1/2$, a minimum value of 0.98605 for $p_v \approx 0.71$, 
   and 1.0 for $p_v=1$. This variation does not show any appreciable finite-size effects. We observed also
   that the exponent $\nu_j$ that characterizes the fluctuation of the jamming state coverage remains 
   consistently the same (within error bars) for all $1/2 \le p_v < 1$. It may be noted that for the RSA without 
   relaxation, $p_j(p_v)$ monotonically decreases from 0.9068 for $p_v=1/2$ to $1 - e^{-2} \approx 0.8647$ for 
   $p_v=1$~\cite{Krapivsky}.

\begin{table}[b]
\caption{Our numerical estimates of the percolation threshold $p_c(p_v)$ in the thermodynamic limit $L \to \infty$, 
   for different values of the selection probability $p_v$ of vertically oriented dimers for RSA with and without
   relaxation. Every reported value has an error bar not more than $2$ in the last digit.
}
\begin{tabular*}{\linewidth}{c@{\extracolsep{\fill}}cr}
\hline \hline \vspace{-0.27cm} \\
 & \multicolumn{2}{c}{$p_c(p_v)$} \\ \vspace{-0.27cm} \\
\cline{2-3} \vspace{-0.27cm} \\
   $p_v$  & RSA with relaxation      & RSA without relaxation    \\ \vspace{-0.27cm} \\ \hline \hline
   0.50   & 0.5140    & 0.5619 \\
   0.58   & 0.5150    & 0.5624 \\
   0.66   & 0.5181    & 0.5640 \\
   0.74   & 0.5232    & 0.5668 \\
   0.82   & 0.5306    & 0.5708 \\
   0.90   & 0.5407    & 0.5764 \\
   0.98   & 0.5539    & 0.5840 \\
   1.00   & 0.5578    & 0.5862 \\ \hline \hline 
\end{tabular*}
\label{TAB01}
\end{table}

      The effect of anisotropy on the percolation threshold has also been studied. For a given value of the 
   anisotropy parameter $p_v$, the asymptotic value of the percolation threshold $p_c(p_v)$ has been 
   determined using the extrapolation method described before. The deviation of $p_c(p_v) - p_c(1/2)$ from 
   the isotropic case has been observed to follow a power law with $p_v - 1/2$ (Fig.\ \ref{fig:pcpv}). On a 
   double logarithmic scale the data points fit with an exponent $= 2.06(6)$. Therefore, we conjecture that 
   $p_c(p_v) - p_c(1/2) \sim (p_v - 1/2)^2$. Our simulation results also predict that this behavior holds 
   for the RSA without relaxation (Fig.\ \ref{fig:pcpv}). In Table \ref{TAB01}, the values of $p_c(p_v)$ for 
   a few values of $p_v$ are listed for RSA with and without relaxation.

      The measured values of the critical exponents $\nu, \gamma, d_f$ and $d_l$ in the entire range of $p_v$ 
   have been found to be consistent within error bars with their respective values for random percolation in 
   two dimensions.

\begin{figure}[t]
\begin{center}
\includegraphics[width=\linewidth]{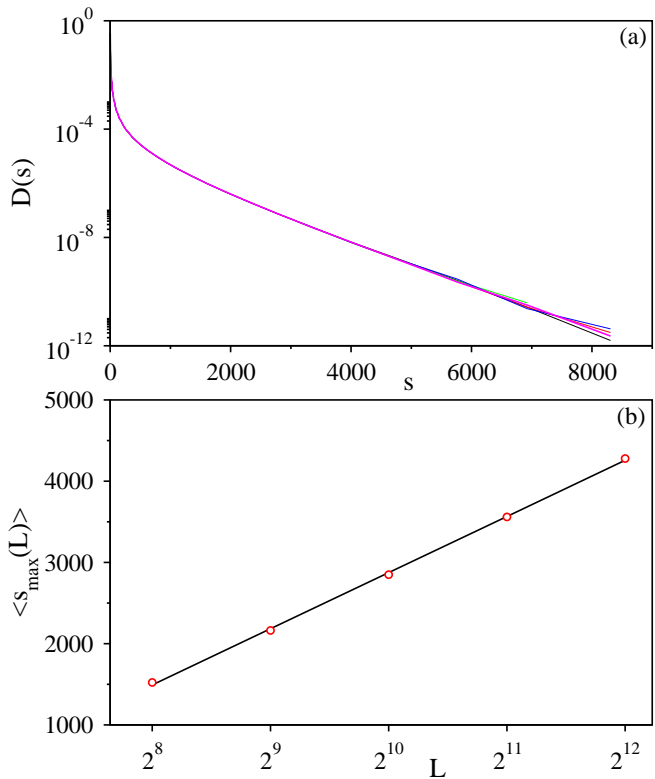}
\end{center}
\caption{Right at the jamming state for the anisotropy parameter $p_v=1/2$, (a) the binned data for cluster 
   size distribution $D(s)$ of the vertically oriented dimers has been exhibited on a semilog scale for $L$ 
   = 256 (black), 512 (red), 1024 (green), 2048 (blue), 4096 (magenta); (b) the average size of the largest 
   cluster $\langle s_{max}(L) \rangle$ for the same values of $L$ has been plotted against $L$ on a lin-log 
   scale. The data points fit considerably well with a straight line indicating the logarithmic growth of the 
   largest cluster. The results are averages over samples ranging from $6.1 \times 10^6$ for $L=256$ to 6300 
   for $L=4096$.
}
\label{cluspj}
\end{figure}

\subsection{Percolation through the sites occupied by similarly oriented dimers in the jamming state}

      Let us now distinguish the clusters of adsorbed dimers by the orientation of the corresponding dimers 
   in the jamming state. The size $s$ of a cluster is the number of sites occupied by the cluster. It is well 
   known that for RSA without relaxation with $p_v=1/2$, the largest among all clusters does not form a 
   spanning path between two opposite boundaries of the lattice~\cite{Lebovka}. As, our model with relaxation 
   dynamics enables more surface coverage, we thus address the question on whether such a spanning cluster 
   appears with relaxation. Identifying different clusters using the burning algorithm~\cite{Stauffer} and 
   using many independent runs, we find that the cluster size distribution $D(s)$ follows an exponential 
   distribution (Fig.\ \ref{cluspj}(a)). Further, the average size of the largest cluster $\langle s_{max}(L) 
   \rangle$ is observed to grow logarithmically with the size of the system (Fig.\ \ref{cluspj}(b)). These 
   results indicate clearly that for $p_v=1/2$, there exist no such spanning cluster and therefore, the system 
   remains in the sub-critical phase of the percolation transition, when clusters are distinguished by the 
   orientation of the dimers in the jamming state. However, $\langle s_{max}(L) \rangle$ for the RSA with 
   relaxation is higher in comparison to the RSA without relaxation and we see that the ratio between them 
   asymptotically approaches to $\approx$ 2.23.

      By increasing the value of $p_v$ from $1/2$, the $\langle s_{max}(p_v) \rangle$ monotonically increases 
   and at a critical value of $p_v=p_{vc}$, the largest cluster first spans the system and percolation of 
   equal-oriented dimers occurs. In numerical simulations, imposing periodic boundary conditions along the 
   horizontal direction, global connectivity along the vertical direction is checked through the neighboring
   sites occupied by vertically oriented dimers.

\begin{figure}[t]
\begin{center}
\begin{tabular}{c}
\includegraphics[width=\linewidth]{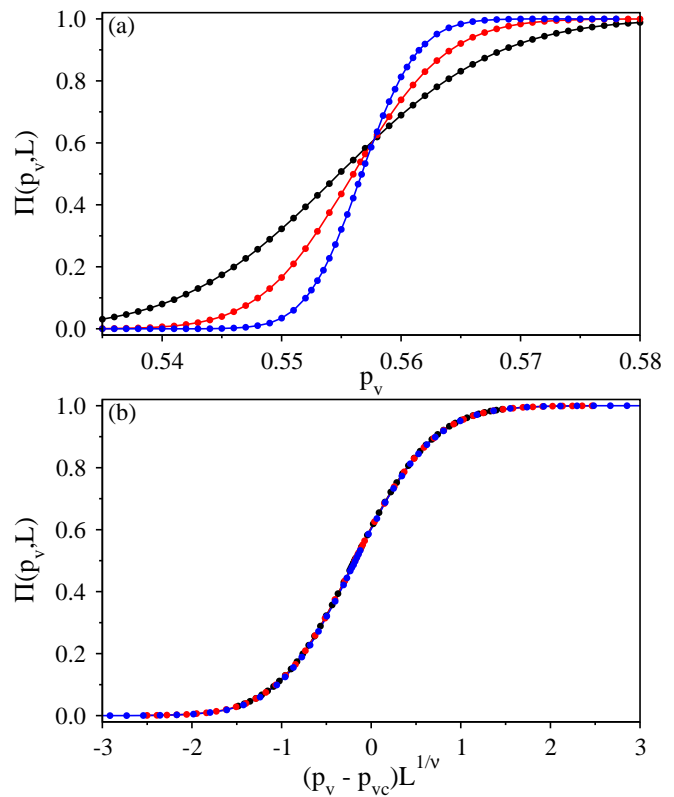}
\end{tabular}
\end{center}
\caption{(a) For $L$ = 256 (black), 512 (red) and 1024 (blue) the percolation probability $\Pi(p_v,L)$ has been 
   plotted with the probability of selection $p_v$ of the vertically oriented dimers. (b) Scaling plot of the 
   same data as in (a). A plot of $\Pi(p_v,L)$ against $(p_v - p_{vc})L^{1/\nu}$ using $p_{vc} = 0.5577(5)$ and 
   $1/\nu=0.754(5)$ exhibits a nice data collapse. The data points are averages over samples ranging from (at 
   least) $2.4 \times 10^6$ for $L=256$ to $10^5$ for $L=1024$.
}
\label{percprob}
\end{figure}

      Tuning the value of $p_v$ and averaging over different uncorrelated jamming state configurations for 
   each $p_v$, we plot the percolation probability $\Pi(p_v,L)$ in Fig.\ \ref{percprob}(a) for three 
   different values of the surface sizes. The curves become more and more sharp as $L$ is increased. All 
   these curves intersect approximately at the same point $[p_{vc}, \Pi(p_{vc})]$ with $p_{vc} \approx 
   0.5577$ and $\Pi(p_{vc}) \approx 0.61$, which is slightly lower than the value $0.636454001$~\cite{Ziff} 
   obtained using Cardy's formula for cylindrical geometry~\cite{Cardy}. Figure \ref{percprob}(b), exhibits 
   a scaling plot of $\Pi(p_v,L)$ against $(p_v - p_{vc})L^{1/\nu}$. The best data collapse for all three 
   curves corresponds to $p_{vc}=0.5577(5)$ and $1/\nu=0.754(5)$, implying a finite-size scaling form
\begin{equation}
   \Pi(p_v,L) = {\cal F}[(p_v - p_{vc})L^{1/\nu}].
\end{equation}
   Similarly, scaling analyses have been performed for the order parameter $\Omega(p_v,L)=\langle s_{max}(p_v,L) 
   \rangle / L^2$ and susceptibility, defined by the fluctuation of the order parameter as $\chi(p_v,L) = \langle 
   \Omega(p_v,L)^2 \rangle - \langle \Omega(p_v,L) \rangle^2$. With a finite-size scaling analysis (not shown) we
   also find that the scaling exponents, $\beta$ and $\gamma$ follow within error bars the hyperscaling relation 
   $2\beta/\nu + \gamma/\nu = 2$ in two dimensions~\cite{Stauffer}.

\begin{figure}[t]
\begin{center}
\begin{tabular}{c}
\includegraphics[width=\linewidth]{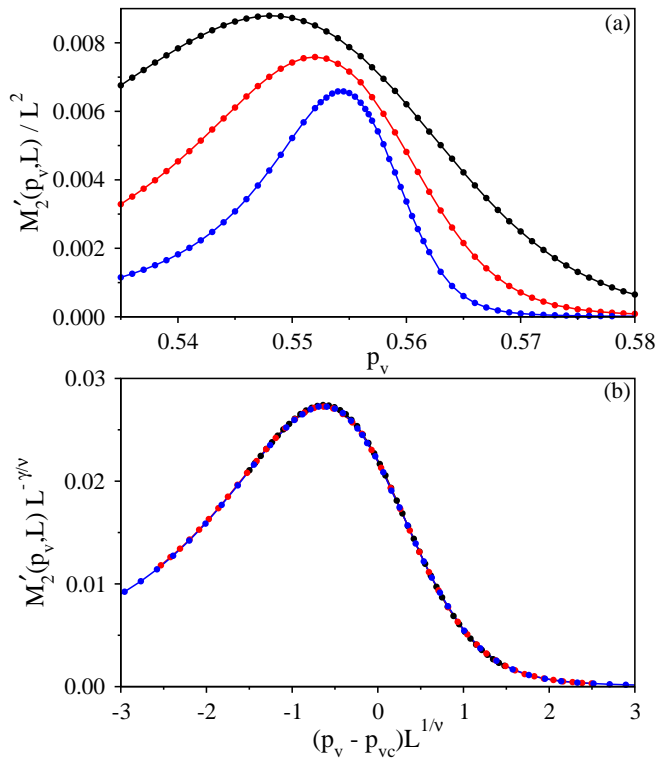}
\end{tabular}
\end{center}
\caption{(a) For $L$ = 256 (black), 512 (red) and 1024 (blue) the scaled second moment $M_2^{\prime}(p_v,L)/L^2$
   has been plotted against the selection probability $p_v$ of the vertically oriented dimers. (b) Finite-size 
   scaling of the same data as in (a). Plot of the re-scaled second moment $M_2^{\prime}(p_v,L)L^{-\gamma/\nu}$ 
   with the scaling variable $(p_v - p_{vc})L^{1/\nu}$ using $p_{vc} = 0.5577(5)$, $1/\nu=0.754(5)$ and 
   $\gamma/\nu=1.795(5)$ exhibits a nice data collapse. The data points are averages over samples ranging from 
   (at least) $2.4 \times 10^6$ for $L=256$ to $10^5$ for $L=1024$.
}
\label{moment}
\end{figure}

      The second moment of the cluster size distribution $M_2^\prime$ is defined as $M_2^\prime=\sum_k s_k^2/L^2 
   - \langle s_{max} \rangle/L^2$ where, $s_k$ being the size of the cluster $k$. In Fig.\ \ref{moment}(a), the 
   behavior of $M_2^\prime(p_v,L)$ has been shown for same three system sizes. By suitably scaling the abscissa 
   and ordinate when the same data are re-plotted, an excellent data collapse is observed using $p_{vc}=0.5577(5)$, 
   $1/\nu=0.754(5)$ and $\gamma/\nu=1.795(5)$, indicating a scaling form
\begin{equation}
   M_2^\prime(p_v,L) = L^{\gamma/\nu}{\cal G}[(p_v - p_{vc})L^{1/\nu}].
\end{equation}
   The same set of scaling analyses have been performed for RSA without relaxation, and we obtain 
   $p_{vc}=0.6056(5)$.

\section{Final remarks}

      We introduce a model of adsorption of dimers on a two dimensional surface, with relaxation. 
   The dimers are sequentially and irreversibly adsorbed on a square lattice at random locations 
   by following a set of predefined conditions. Most importantly, a relaxation dynamics is involved 
   with the adsorption process. When a newly deposited dimer partially overlaps with a previously 
   adsorbed dimer, a sequence of dimer displacements may occur to accommodate the new dimer. Every 
   adsorption followed by the relaxation dynamics includes a pair of new occupied sites separated 
   by a distance larger than unity and therefore, setting in spatial correlations. The effect of 
   the relaxation dynamics and anisotropy in the orientation of the adsorbed dimers on the jamming 
   state and percolation transition have been investigated in detail. 

      The percolation transition for the isotropic case occurs at a critical density of occupied sites 
   $p_c=0.5140(1)$. The increase of anisotropy, $p_v$, of the occurrence of vertical dimers results in 
   an increase of the percolation threshold. In comparison to the Random Sequential Adsorption (RSA) model 
   without relaxation, the percolation threshold in the entire range of $p_v$ is much lower for our model
   with relaxation. Using extensive numerical simulations and measuring different critical exponents 
   associated with the transition lead us to conclude that, despite the developed spatial correlations, 
   the percolation transition always fall into the random percolation universality class.

      The jamming state coverage is higher for RSA with relaxation than without relaxation. A non-monotonic 
   variation of jamming state coverage with the strength of anisotropy $p_v$ has been observed for RSA with
   relaxation. Further, separating out the vertically oriented dimers from the horizontal ones in the jamming 
   state, a percolation transition through the cluster of sites occupied by vertically oriented dimers is 
   observed when the control parameter $p_v$ is tuned to the critical value $p_{vc}=0.5577(5)$. Also here, 
   the directionality does not affect the critical (universal) properties of the percolation transition. 

\begin{acknowledgments}
 
      S.K. and N.A.M.A. acknowledge financial support from the Portuguese Foundation for Science and Technology 
   (FCT) under the contract no. UID/FIS/00618/2013. S.K. acknowledges senior research fellowship, provided by
   S. N. Bose National Centre for Basic Sciences, Kolkata, India.
 
\end{acknowledgments}

\end {document}